\documentclass[%
 reprint,
superscriptaddress,
showpacs,
 amsmath,amssymb,
 aps,
]{revtex4-1}
\usepackage{amsmath}
\usepackage{amsfonts}
\usepackage{graphicx}
\usepackage{dcolumn}
\usepackage{amsbsy}
\usepackage{bm}
\usepackage{amsthm}
\usepackage[caption=false]{subfig}
\usepackage{tabularx}
\usepackage{array}
\usepackage{natbib}

\begin{document}

\title{Correlated coherence as resource of remote state preparation and quantum teleportation}

\author{Shenghang Li}
 \affiliation{Zhejiang Institute of Modern Physics, Zhejiang University, Hangzhou 310027, China}
\author{Yi-Xin Chen}
 \email{yixinchenzimp@zju.edu.cn}
 \affiliation{Zhejiang Institute of Modern Physics, Zhejiang University, Hangzhou 310027, China}
\date{\today}

\begin{abstract}
Quantum discord and quantum entanglement are resources in some quantum information processing (QIP) models. However, in recent years, the evidence that separable states or classically correlated states can also accomplish QIP is demonstrated. It provides a useful tool since such states are easier to prepare. Quantum coherence is a measure of non-classical correlation, containing entanglement and discord as a subset. Nowadays, it is of interest whether quantum coherence can act as a resource in QIP independently or not, without the help from quantum discord or entanglement. In this paper, we show that quantum correlated coherence(a measure of coherence with local parts removed) is also a kind of quantum resource. It is the sufficient and necessary resource for quantum remote state preparation and quantum teleportation. 
\end{abstract}
\maketitle

\section{Introduction}
\label{seci}
Different from classical mechanics, quantum mechanics admit a superposition containing more information. Quantum correlation and quantum coherence represent structural features of quantum superposition\cite{nielsen2002quantum}. While quantum correlation is defined in a system which involved at least two parties, quantum coherence can be defined for the single system \cite{baumgratz2014quantifying}. Also, for a quantum system, they also demonstrate different characteristics. Although the above two differences, arising from superposition, the interrelation between quantum correlation and quantum coherence is essential to study. Quantum discord and quantum entanglement are representatives of quantum correlation because of their resource power. The interrelations between coherence and entanglement \cite{tan2016unified,streltsov2015measuring,mondal2017nonlocal}, coherence and discord \cite{Yao2015Quantum,Qi2016Measuring,Xi2015Quantum,Ma2016Converting} have been discussed.

Coherence could provide bound of entanglement and discord \cite{tan2016unified,streltsov2015measuring,mondal2017nonlocal, Yao2015Quantum} and coherence can be transformed to discord. For example, based on the definition of entanglement $E$, discord $D$ and coherence $C$, we have an inequality equation that $C\ge D\ge E$ for a given quantum state under relative entropy measurement \cite{Yao2015Quantum}. If a coherent state S is coupled to an incoherent ancilla state A, whose initial state is the vacuum state$|0^A\rangle$, the generated entanglement $E_{SA}$ satisfies $E_{SA}\le C_S$ \cite{streltsov2015measuring}. Quantum discord is equal to basis-free coherence $D(\rho)=\min\limits_{\vec{U}}C(\vec{U}\rho\vec{U}^\dagger)$ \cite{Yao2015Quantum}.

Quantum correlation and quantum coherence are optimal quantum information carriers. They are believed as the cornerstones of quantum communication and quantum computation, which is hardly accomplished by classical state. In previous researches, it is discovered that quantum correlation is a key resource of quantum information process. Quantum entanglement is a resource for non-local tasks \cite{Ekert1991Quantum,bennett1993teleporting} which could not be finished by separable state and local operations and classical communications(LOCC) \cite{werner1989quantum}. However, in some tasks, quantum entanglement plays little role \cite{knill1998power,braunstein1999separability,lanyon2008experimental}. Other correlations should be more influential in such tasks. It is of interest to discover new correlations. Additionally, the creation and manipulation of entanglement are difficult, so in experiments, the systems are limited to small-scale systems \cite{monz201114,wang2016experimental,barends2014superconducting}. It is another motivation to find new correlations. Later, in some situations that entanglement is not efficient, discord works as a necessary resource.  \cite{horodecki2005local,datta2008quantum,madhok2013quantum,dakic2012quantum}, despite there are some disputes \cite{modi2012classical,datta2012quantum,ferraro2012nonclassicality,giorgi2013quantum}. Recently, as another embodiment of the superposition principle, quantum coherence has been shown as important physical resource in quantum thermodynamics \cite{lostaglio2015description,lostaglio2015quantum,narasimhachar2015low,gour2015resource}, reference frames \cite{bartlett2007reference}, quantum biology \cite{lambert2013quantum}, and nanoscale physics \cite{vazquez2012probing,karlstrom2011increasing}. The study of the similarity between the resource theory of coherence and that of entanglement \cite{horodecki2009quantum}, the interrelation between entanglement, discord, and coherence have been done. Whether coherence except entanglement or discord acts as a resource or not is of interest. Besides, another motivation is that coherent states are easier to prepare 
than entangled or discordant states. In this paper, we take correlated coherence in consideration and discuss its power in remote state preparation (RSP) protocol and quantum teleportation, where the entanglement and discord act as resource and classical correlated states are also useful.

This paper is organized as follows: in Sec. \uppercase\expandafter{\romannumeral2}, we introduce the definitions and resource power of quantum entanglement, quantum discord, and quantum coherence. In Sec.\uppercase\expandafter{\romannumeral3}, we provide an optimal representation of correlated coherence for the bipartite system. In Sec. \uppercase\expandafter{\romannumeral4}, we introduce the RSP protocol, then provide the relation between correlated coherence and average payoff of the RSP protocol, to prove correlated coherence acting as a resource. In Sec. \uppercase\expandafter{\romannumeral5}, we introduce the average fidelity of quantum teleportation, then provide the relation between the average fidelity and correlated coherence. Finally, we show a conclusion in Sec. \uppercase\expandafter{\romannumeral6}

\section{Quantum correlation and quantum coherence}
Quantum entanglement is one of the most important measures of quantum correlation. As mentioned before, quantum entanglement is discovered in composite systems. For a multipartite system consisting of $n$ subsystems, the Hilbert space is the tensor product of the $n$ subsystem spaces. Then, for the pure state, the quantum state which is a product of n states of individual subsystem is named as separable state
$
|\psi\rangle=|\psi_1\rangle\otimes|\psi_2\rangle\otimes...\otimes|\psi_n\rangle
$.
For mixed state, not only the product states but also the convex combination of product states are separable states. The separable state is defined as the superposition of the product of n states of individual system 
$\rho=\sum_ip_i\rho_1^i\otimes...\otimes\rho_n^i$. 
Such separable state can be simulated by preparing it from classical correlation \cite{werner1989quantum}. However, not all of the quantum states are separable. The state cannot be simulated by classical correlation are named entangled state. Then, quantum entanglement is defined as the minimum distance between a given state $\rho_{AB}$ and a separable state: $E(\rho)=\min\limits_{\delta\in S}\mathbb{D}(\rho||\delta)$, where $S$ stand for the set of separable state. 

Quantum entanglement is a crucial resource in some non-local tasks. In RSP protocol and quantum teleportation, maximally entangled states are the most efficient states because the fidelity is equal to 1. However, the evidence that another kind of quantum resource exists is provided, such as accomplishing RSP protocol and quantum teleportation with separable states. Later, quantum discord is shown as another critical resource. 

For a bipartite system, quantum discord is defined in terms of quantum mutual information(QMI) firstly. Mutual information has two expressions: $\mathbb{I}(\rho_{AB})=S(\rho_A)+S(\rho_B)-S(\rho_{AB})$, with $S(\rho_X)=-tr(\rho_Xlog_2\rho_X)$ being Von Neumann entropy, and  $J(\rho_{AB})=S(\rho_B)-S(\rho_B|\{\Pi_k^A\})$. These two expressions are equal in classical case. However, in non-classical case, they are probably not identical. Quantum discord is defined as the minimum difference between these two expressions $D=\min\limits_{\{\Pi_k^A\}}\mathbb{I}(\rho_{AB})-J(\rho_{AB})$. Here, $\{\Pi_k^A\}$ is the set of von Neumann measurements (one-dimensional orthogonal projectors that sum to identity) performed on the party A, with the conditional state $\rho_k=\frac{1}{p_k}(\Pi_k^A\otimes I_B)\rho(\Pi_k^A\otimes I_B)$, probability $p_k=tr(\Pi_k^A\otimes I_B)\rho(\Pi_k^A\otimes I_B)$ and the averaged conditional entropy $S(\rho_B|\{\Pi_k^A\}) = \sum_kp_kS(\rho_k)$. Based on the formula, the intuitive meaning of quantum discord is the minimum loss of correlation due to measurement. Then, analog to the definition of quantum entanglement, a general definition of quantum discord is introduced. Quantum discord is defined as the minimum distance between a given state $\rho_{AB}$ and a classically correlated state: $D(\rho)=\min\limits_{\delta\in CC}\mathbb{D}(\rho||\delta)$, where $CC$ stand for the set of classically correlated state(i.e. $D(\delta)=0$).

From the definition of separable state $S$ and classically correlated state $CC$, the inclusion of these sets emerges $CC\subset S$. It is clear that $D\ge E$. Quantum discord contains more information than entanglement. As mentioned before, quantum entanglement and quantum discord are practical resources in some QIPs. However, it has previously been observed that some quantum states without quantum discord or entanglement partly could accomplish RSP protocol also. We believe that there exists another definition of quantum correlation containing quantum discord as a subset. This definition describes the resource power also. We take coherence in consideration since it contains more information than quantum discord, 


As the above two definitions, quantum coherence has a similar definition. It is defined as the minimum distance between a given state $\rho_{AB}$ and a incoherent state: $C(\rho)=\min\limits_{\delta\in \mathcal{I}}\mathbb{D}(\rho||\delta)$, where $\mathcal{I}$ stand for the set of incoherent state($\delta = \sum_{i=1}^d p_i\delta_{i,i}|i\rangle\langle i|$).

To be a proper coherence measure, the distance measure should satisfy some conditions introduced by Baumgratz \cite{baumgratz2014quantifying}. The conditions are follows:

($C1$) Negativity 
\\
\centerline{$C(\rho)\ge0$, and $C(\rho)=0$ iff $\rho\in{\mathcal{I}}$.}

($C2a$) Monotonicity under incoherent completely positive and trace preserving maps(ICPTP), i.e., 
\\
\centerline{$C(\rho)\ge C(\mathbb{\phi}_{ICPTP}(\rho))$.} 
\\Here, the ICPTP operations $\phi_{ICPTP}(\rho)$ act as $\phi_{ICPTP}(\hat{\rho})=\sum_i\hat{K}_i\hat{\rho}\hat{K}_i^\dagger$. The Kraus operators $\hat{K}_i$ are all of the same dimension and satisfy $\hat{K}_i\delta\hat{K}_i^\dagger/p_i\in{\mathcal{I}}$ for arbitrary $\delta\in{\mathcal{I}}$, with $p_i=tr(\hat{K}_i\rho\hat{K}_i^\dagger)$ being the probability of result $i$.

($C2b$) Monotonicity under selective measurements on average, i.e. 
\\\centerline{$C(\rho)\ge \sum_i p_iC(\rho_i)$.}
\\ Here, the incoherent operations with subselection also need $\hat{K}_i\delta\hat{K}_i^\dagger/p_i\in{\mathcal{I}}$ for arbitrary $\delta\in{\mathcal{I}}$. However, the dimension of $\hat{K}_i$ may be different.

($C3$) Convexity: coherence doesn't increase under mixing of quantum states, i.e. 
\\
\centerline{$\sum_i p_i C(\rho_i)\ge C(\sum_i p_i \rho_i)$,}
p is the probability satisfying any $p_i\ge 0$ and $\sum_i p_i = 1$

Particularly, coherence measures satisfying ($C2b$) and ($C3$) always satisfy the condition ($C2a$), since $C(\rho)\ge \sum_i p_iC(\rho_i)\ge C(\sum_i p_i \rho_i)=C(\mathbb{\phi}_{ICPTP}(\rho))$

($C4$) Additivity:  
\\\centerline{$C(p_1 \rho_1 \oplus p_2 \rho_2) = p_1C(\rho_1)+p_2C(\rho_2)$}
($C2b$) and ($C3$) could be replaced by condition ($C4$).
\\\indent
So far, there are many coherence measures proposed. Some of them satisfy the conditions above and some of them not. Two proper measures of coherence are follows: 

(a) The relative entropy of coherence: 
\\\centerline{$C_r(\rho)=S(\rho_{diag})-S(\rho)$,} 
where $\rho_{diag}$ is the diagonal part of $\rho$: $\rho_{diag}=\sum_i|i\rangle \langle i|\rho |i\rangle \langle i|$.

(b) $l_1$-norm of coherence: $C_{l_1}=\sum_{i\neq j}|\rho_{i,j}|$





Similar as the relation between discord and entanglement, from the definition of separable state $S$, classically correlated state $CC$, and incoherent state $\mathcal{I}$, the inclusion of these sets emerges as $\mathcal{I}\subset CC\subset S$. So, it is clear that $C\ge D\ge E$. Coherence contains more information than entanglement and quantum discord. However, while quantum discord and entanglement are defined in a system which involved at least two parties, quantum coherence can be defined in a single system. Thus, for bipartite systems, coherence contains both local correlation and non-local correlation. We agree with the opinion that the non-local correlation is helpful in QIPs. We introduce correlated coherence as a tool to remove the local part of coherence. Then, the non-local part of coherence, a global quantum correlation, is provided and useful to discuss the resource power of a state. 

\section{Correlated coherence}
Correlated coherence is defined in \cite{tan2016unified}:
\begin{eqnarray}
    C_{cor}(\rho)=C(\rho)-C(\rho_A)-C(\rho_B)
\end{eqnarray}
When we adopt the relative entropy of coherence or $l_1$-norm of coherence, the result is complicated and hard to analyze. Among the various measures of coherence, we pick up a suitable measure of coherence, the $l_2$-norm coherence.

A rigorous framework for the quantification of coherence is introduced before \cite{baumgratz2014quantifying}. In this article, measures induced by $l_p(p>1)$ matrix norms do not serve as proper coherence monotones. However, a new viewpoint of the incoherent state is mentioned recently \cite{yu2016total}. To overcome the basis dependence of quantum coherence, we use the maximally mixed state as incoherent state \cite{radhakrishnan2018basis,yu2016total,yao2016frobenius,hu2017maximum,zhang2017classical}. Compared with the basis-dependent incoherent state, $\delta = \sum_{i=1}^d p_i\delta_{i,i}|i\rangle\langle i|$, the maximally mixed state $\zeta=I_d/d$ allows a new definition of coherence that is invariant under unitary transformation. Under this circumstance, induced by $l_2$ matrix norms, 
\begin{equation}
    C_{l_2}=\sum_{i,j=1}^d|\rho_{i,j}-\zeta_{i,j}|^2=tr\rho^2-1/d
\end{equation} 
is a proper measurement of coherence \cite{yu2016total}.

For simplification, we define a coherence measure $C_2=d\times C_{l_2}=d\times tr\rho^2-1$. For a d-dimension Hilbert space, the maximum coherence $(C_2)_{max}=d-1$ and the minimum coherence $(C_2)_{min}=0$. For bipartite system, correlated coherence\cite{tan2016unified}

\begin{eqnarray}
    C_c(\rho)=C_2(\rho)-C_2(\rho_A)-C_2(\rho_B)
\end{eqnarray}
with $\rho_A=tr_B(\rho)$ and $\rho_B=tr_A(\rho)$, could be resource for remote state preparation and quantum teleportation.

\section{Remote state preparation}
In this section, we shall show the relation between correlated coherence and spherical average payoff in RSP protocol $
P=\frac{1}{3}C_c(\rho),
$
 as support of quantum correlated coherence acting as a resource.

Remote state preparation (RSP) is a quantum protocol which could prepare a pure state remotely by local operators and classical communication (LOCC). This protocol is different from quantum state teleportation that Alice knows what state needs transmitting to Bob in advance. 

To accomplish the RSP protocol \cite{dakic2012quantum}, Alice and Bob share a two-qubit state $\rho$, which can be represented as 
\begin{equation}
      \rho=\frac{1}{4}(I_{AB}+\sum_{i=1}^3\sigma_i a_i\otimes I_B+\sum_{j=1}^3 I_A\otimes\sigma_j b_j+\sum_{i,j=1}^3\sigma_i\otimes\sigma_jE_{ij})
\end{equation}

Here, $\sigma_i$ ($i=1,2,3$) are the Pauli matrices, $\vec a = (a_1,a_2,a_3)$ is the Bloch vector and $\{E_{ij}\}=Tr[\rho(\sigma_i\otimes\sigma_j)]$ denotes the correlation matrix.

If Alice wants to remotely prepare a quantum pure state $\vec{s}$ perpendicular to a direction $\vec{\beta}$(which is announced to Bob before), she measures along a direction $\vec{\alpha}$ and obtains $\alpha=1$ or $\alpha=-1$ with the probability $\mathbb{P}(\alpha)=(1+\alpha\vec{\alpha}\cdotp\vec{a})/2$. Then, she sends her result $\alpha=\pm1$ to Bob via a classical channel as a correction to the state. After this measurement, Bob will obtain his state $\vec{b_\alpha}=\frac{\vec{b}+\alpha E^T\vec{\alpha}}{1+\alpha\vec{\alpha} \cdotp\vec{a}}$
with the probability $\mathbb P(\alpha)=(1+\alpha\vec{\alpha}\cdotp\vec{a})/2$. Here the $E_{ij}$ is the correlation tensor in (4) and $T$ means transpose. With Alice's result $\alpha$=1 or -1, Bob applies a conditional rotation on his state. The state on Bob's side turns to be $\vec{r}=\mathbb P(+1)\vec{b_+}+\mathbb P(-1)R_\pi \vec{b_-}$ with $R_\pi=-1$.  The fidelity of this protocol is 
\begin{eqnarray}
F=\frac{1}{2}(1+\vec{r}\cdotp\vec{s})=\frac{1}{2}(1+\vec{\alpha} \overset{\text{\tiny$\leftrightarrow$}}{E} \vec{s})
\end{eqnarray}
and the payoff function is 
\begin{eqnarray}
P=(2F-1)^2=(\vec{\alpha} \overset{\text{\tiny$\leftrightarrow$}}{E} \vec{s})^2
\end{eqnarray}

Here, the payoff $P$ is directly related to the fidelity $F$ as $P=(2F-1)^2$. For a given target state $\vec s$ and direction $\vec\beta$, Alice could optimize the fidelity by the choice of the local measurement direction $\vec\alpha$. For the fidelity $F=\frac{1}{2}(1+\vec{r}\cdotp\vec{s})$, we can rotate $\vec{r}$ to maximize $F$ to make sure that $F\ge \frac{1}{2}$. In the RSP protocol, the maximation is achieved by the selection of $\vec{\alpha}$, where $\vec{\alpha}$ is parallel to $\overset{\text{\tiny$\leftrightarrow$}}{E} \vec{s}$ and $|\vec{\alpha}| = 1$.

Since $F=\frac{1}{2}(1+\vec{r}\cdotp\vec{s})\ge\frac{1}{2}$ and $P=(2F-1)^2$, payoff $P$ and fidelity $F$ are optimized meanwhile. The chosen $\vec\alpha$ induces the maximum $P$ as $P = \sum_{i=1}^3(\sum_{j=1}^3 E_{i,j}s_j)^2$. When $\rho$ is a totally mixed state, the payoff satisfies $P=0$. While $\rho$ is a maximally entangled state, we have $P=1$. Based on these properties, the payoff function can be used to determine the efficiency of RSP protocol. It is also helpful in discussing resource power. From the Eq.(6), payoff function is regarded as quadratic fideilty \cite{kanjilal2018remote}.

It is shown in \cite{dakic2012quantum} that quantum geometric discord could be resource for RSP, while the target state $\vec{s}$ has a zero component in the z-direction of Alice (i.e., $\vec{\beta}$ corresponds to z) and the Bloch vector $\vec a$ is parallel to the eigenvector $\vec{v_1}$ correspond to the largest eigenvalue of $E^TE$. 

The average payoff $P$ over the all possible vector $\vec{s}=(cos\theta, sin\theta, 0)$ is
\begin{equation}
\begin{aligned}
    \langle P_{opt}\rangle 
      &= \frac{1}{2\pi}\int_0^{2\pi}d\theta\sum_{k=1}^3(E_{k1}cos\theta+E_{k2}sin\theta)^2
    \\&= \frac{1}{2}(E_{11}^2+E_{21}^2+E_{31}^2+E_{12}^2+E_{22}^2+E_{32}^2).
\end{aligned}
\end{equation} 
Here, since $\vec\beta$ is selected arbitrarily, the authors of \cite{dakic2012quantum} analyze the worst case over $\vec\beta$ to determine the fidelity of RSP protocol for a given state. The RSP quadratic fidelity\cite{dakic2012quantum,kanjilal2018remote} is defined as the minimal average payoff  $P_{min}=\min\limits_{\vec{\beta}}\langle P_{opt}\rangle$. They pick the lowest payoff function to determine the efficiency of the RSP protocol for a given state $\rho$. The lowest payoff is the minimum $P_{opt}$ over all $\vec\beta$. Since 
\begin{equation}
||E||^2=\sum_{n=1}^3E_n^2=\sum_{i,j=1}^3E_{i,j}^2
\end{equation}
is a constant independent of the choice of $\vec\beta$, Eq.(7) can be represented as 

\begin{eqnarray}
\langle P_{opt} \rangle =\frac{1}{2}[\sum_{i,j=1}^3E_{i,j}^2-(E_{13}^2+E_{23}^2+E_{33}^2)]
\end{eqnarray}

Minimization of (9)
is equivalent to maximize $E_{13}^2+E_{23}^2+E_{33}^2$. Then, from the relation that $E_{13}^2+E_{23}^2+E_{33}^2=\vec{\beta}^T(E^TE)\vec{\beta}$, the maximum of $E_{13}^2+E_{23}^2+E_{33}^2$ is the largest eigenvalues of $E^TE$. 
The RSP quadratic fidelity $P_{min}=\min\limits_{\vec{\beta}}\langle P_{opt}\rangle=\frac{1}{2}(E_2^2+E_3^2)$, where $E_i^2$s are the eigenvalues of $E^TE$ in decreasing order. 

Besides, the (normalized) geometric measure of quantum discord is 
\begin{equation}
D^2 =\frac{1}{2}(||\vec{a}||^2+||E||^2-k_{max})
\end{equation}
Here, $\vec{a}$ is the local Bloch vector, $k_{max}$ is the largest eigenvalue of $K=\vec{a}\vec{a}^T+EE^T$. Here we focus on the class of state, whose bloch vector $\vec a$ is parallel to the eigenvector $\vec{v_1}$ correspond to the largest eigenvalue of $E^TE$. For such states, in the eigenbasis of $E^TE$, the matrix $E^TE$ is represented as $E^TE=diag[E_1^2,E_2^2,E_3^2]$, and the local bloch vector is written as $\vec a=(\kappa,0,0)$. Thus, the matrix K has the form $K = diag[E_1^2+\kappa^2,E_2^2,E_3^2]$ with the largest eigenvalue $E_1^2+\kappa^2$.
\begin{equation}
\begin{aligned}
    D^2 &=\frac{1}{2}(||\vec{a}||^2+||E||^2-k_{max}) 
    \\&= \frac{1}{2}(|\kappa|^2+E_1^2+E_2^2+E_3^2-(E_1^2+\kappa^2)) 
    \\&=\frac{1}{2}(E_2^2+E_3^2).
\end{aligned}
\end{equation} 

The relation $D^2 = P_{min}$ is provided.

However, this property is not universal since two conditions have to be satisfied. The target state has a zero component in the z-direction of Alice, and the eigenvector corresponding to $E_1^2$ is parallel to the direction of the local Bloch vector $\vec a$. Without these conditions, we could find a classically correlated state to achieve non-zero payoff for RSP protocol and a discordant state to achieve zero payoff \cite{giorgi2013quantum}. So, we aim to find other correlation which is more relevant to RSP protocol payoff without these conditions. In the paragraphs below, we show that correlated coherence of the state is a more proper correlation.

Now, we focus on a universal situation that source state $\rho$ is arbitrary, while the Bloch vecctor $\vec a$ is restricted before, and the target state is arbitrary on the surface of Bloch sphere, which is limited in xy-plane before. And we analyse spherical average payoff instead of the circular payoff.
\\\indent For a target state $\vec{s}=(sin\theta cos\phi, sin\theta sin\phi, cos\theta)$, $P_s=\sum_{k=1}^3(E_{k1}sin\theta cos\phi+E_{k2}sin\theta sin\phi+E_{k3}cos\theta)^2$. And the average payoff is

\begin{equation}
    P=
    \frac{\int_{\theta=0}^{\pi}sin\theta\int_{\phi=0}^{2\pi}P_s d\theta d\phi}{\int_{\theta=0}^{\pi}sin\theta\int_{\phi=0}^{2\pi}d\theta d\phi} = 
    \frac{1}{3} \sum_{i,j=1}^3 E_{i,j}^2
\end{equation}

From equation (4), it is derived that $E_{i,j} = Tr[\rho \cdotp(\sigma_i \otimes \sigma_j)]$, and

\begin{equation}
\begin{aligned} 
    \sum_{i,j=1}^3 E_{i,j}^2 
     =&(\rho_{11}^2+\rho_{22}^2+\rho_{33}^2+\rho_{44}^2-2\rho_{11}\rho_{22}-2\rho_{11}\rho_{33}
    \\&+2\rho_{11}\rho_{44}+2\rho_{22}\rho_{33}-2\rho_{22}\rho_{44}-2\rho_{33}\rho_{44})
    \\&+(8\rho_{41}\rho_{14}+8\rho_{23}\rho_{32})
    \\&+(4\rho_{13}\rho_{31}-4\rho_{13}\rho_{42}-4\rho_{24}\rho_{31}+4\rho_{24}\rho_{42})
    \\&+(4\rho_{12}\rho_{21}-4\rho_{12}\rho_{43}-4\rho_{34}\rho_{32}+4\rho_{34}\rho_{43})
    \\&=C_c(\rho)
\end{aligned}
\end{equation}

Now we get a relation that 
\begin{equation}
\begin{aligned} 
    P=\frac{1}{3}C_c(\rho)
\end{aligned}
\end{equation}
 It shows that correlated coherence is a necessary resource for RSP protocol. Particularly, we select a quantum state $\rho_{AB}$, which satisfying $\rho_{AB}=\rho_A\otimes\rho_B$ , $
    \rho_A=
        \begin{pmatrix}
        0.9 & 0\\
        0 & 0.1
        \end{pmatrix}$ and $
    \rho_B=
        \begin{pmatrix}
        0.8 & 0\\
        0 & 0.2
        \end{pmatrix}
$
.While the entanglement $E(\rho)=0$ and the discord $D(\rho)=0$, the correlated coherence $C_c(\rho) = 0.2304$ and the average payoff $P=0.0768=\frac{1}{3}C_c(\rho)$. It describes that a state without entanglement or discord could accomplish RSP protocol, while correlated coherence acts as the resource. It is powerful evidence that correlated coherence acts as a quantum resource different from entanglement or discord. Besides, for any state $\rho$ with zero correlated coherence and arbitrary target state $\vec{s}$, the state on Bob's side is $\vec{r}=0$. The information provided through $\rho$ is as few as that through the maximally mixed state. The states with zero correlated coherence are the most useless in RSP protocol, as useless as the maximally mixed state.

\section{Quantum teleportation}
In this section, we present that a channel state with non-zero correlated coherence and zero discord is efficient for quantum teleportation, while the states with zero correlated coherence are the most useless ones, as the support for quantum correlated coherence acting as a quantum resource. 
 
A standard teleportation scheme is provided in \cite{horodecki1996teleportation}. In this scheme, Alice and Bob share a quantum channel $\rho$, represented by Eq.(4). One of the particles is given to Bob while another one and a third one are given to Alice. If Alice performs the Bell basis states as measurements onto his qubits and Bob can perform arbitrary unitary operations to his qubit, the highest average fidelity is shown as: 
\begin{equation}
    \tilde{F}=\frac{1}{2}(1+\frac{1}{3}Tr(\sqrt {E^TE})). 
\end{equation}
From the scheme and equation above, the worst case of teleportation is $\tilde{F}=\frac{1}{2}$, induced by $||E||=0$. 
\\\indent 
Besides, the relation between the renormalized geometric quantum discord 
    $\tilde{D}_G(\rho)=\frac{1}{3}(||\vec{a}||^2+||E||^2-k_{max})$ 
and the fidelity are shown  in \cite{adhikari2012operational,verstraete2002fidelity,girolami2011interplay}: 
    $\frac{1+\tilde{D}_G(\rho)}{2}\le\tilde{F}\le\frac{2+\sqrt{\tilde{D}_G(\rho)}}{3}$. 
For correlated coherence and the fidelity, although there is not an equation like the formula(14), we also support correlated coherence acts as a resource. The reasons are following: quantum channel state with correlated coherence and zero discord could accomplish this protocol with $\tilde F\ge\frac{1}{2}$, and quantum channel state with zero correlated coherence induces a maximum average fidelity
    $\tilde{F}=\frac{1}{2}$, which is noneffective. 

For detail, if the channel state is represented as 

    $\rho=\begin{pmatrix}
        \frac{1}{4} & -\frac{i}{16} & \frac{1}{16} & -\frac{i}{16} \\
        \frac{i}{16} & \frac{1}{4} & -\frac{i}{16} & \frac{1}{16} \\
        \frac{1}{16} & \frac{i}{16} & \frac{1}{4} & -\frac{i}{16} \\
        \frac{i}{16} & \frac{1}{16} & \frac{i}{16} & \frac{1}{4}
    \end{pmatrix},$

the maximum averaged fidelity of teleportation is $\frac{13}{24}$ while the correlated coherence is $\frac{1}{16}$, discord is 0 and entanglement is 0. 

For arbitrary state containing zero correlated coherence, from Eqs. (8) and (13) it is derived that $C_{c}=E_1^2+E_2^2+E_3^2$. Then we got that $E_1=E_2=E_3=0$, and the fidelity $\tilde{F}=\frac{1}{2}(1+\frac{1}{3}Tr(\sqrt {E^TE}))=\frac{1}{2}[1+\frac{1}{3}(|E_1|+|E_2|+|E_3|)]=\frac{1}{2}$. As mentioned before, $\tilde{F}=\frac{1}{2}$ means that these states are the most invalid states, as useless as the maximally mixed state. Furthermore, any state with non-zero correlated coherence will induce $|E_1|>0$ and $\tilde{F}>\frac{1}{2}$.
\\\indent What needs noticing is that unitary operators never change the spherical average payoff of RSP protocol or the average fidelity of quantum teleportation.  Moreover, the definition of correlated coherence separates the local coherence $C(\rho_{A})$ and $C(\rho_{B})$ from the total coherence $C(\rho_{AB})$.
\section{Conclusion}
In this work, we provide a new definition of quantum correlation based on renormalized $l_2$-norm coherence: correlated coherence. We show that non-zero correlated coherence is the sufficient and necessary resource for RSP protocol and quantum teleportation. Particularly, it is a different kind of resource from quantum discord and quantum entanglement. Quantum states with non-zero correlated coherence, zero discord, and zero entanglement are efficient, while quantum states without correlated coherence are the most useless states.  Correlated coherence also acts as a cornerstone for quantum information process.
\acknowledgements
This work is partially supported by the NNSF of China, Grant No. 11775187.


%

\end{document}